\documentclass[12pt]{article} 

\usepackage{graphicx}
\usepackage{amssymb}
\usepackage{amsmath,amsfonts}
\usepackage{slashed}
\usepackage{enumerate}

\newcommand{\ii} {\text{i}}
\newcommand{\IM} {\text{Im}\,}
\newcommand{\RE} {\text{Re}\,}
\newcommand{\res}{\,\text{res}\,}

\newcommand{\ba}{\begin{eqnarray}}
\newcommand{\ea}{\end{eqnarray}}
\newcommand{\baa}{\begin{eqnarray*}}
\newcommand{\eaa}{\end{eqnarray*}}
\newcommand{\bb}{\begin{thebibliography}}
\newcommand{\eb}{\end{thebibliography}}

\def\bea{\begin{eqnarray}}
\def\eea{\end{eqnarray}}

\numberwithin{equation}{section}

\begin{document}

\normalsize

\begin{center} {\Large \bf Solitons and Normal Random Matrices} \\

\vspace{5mm}

I. M. Loutsenko$^{1}$, V. P. Spiridonov$^{2,3}$ and O. V. Yermolayeva$^{1}$ \\

\vspace{2mm}

${}^1$ Laboratoire de Physique Math\'ematique,
Centre de Recherches Math\'ematiques,\\ Universit\'e de Montr\'eal, Montr\'eal \\

\vspace{2mm}

${}^2$ Laboratory of Theoretical Physics, Joint Institute for Nuclear Research, Dubna

\vspace{2mm}

${}^3$ National Research University ``Higher School of Echonomics'', Moscow

\vspace{5mm}

\begin{abstract}

{

We discuss a general relation between the solitons and statistical mechanics and
show that the partition function of the normal random matrix model can be obtained
from the multi-soliton solutions of the two-dimensional Toda lattice hierarchy in a special limit.

}

\end{abstract}

\end{center}

\section{Introduction}

Present paper is devoted to intersections between the theory of solitons and statistical mechanics of
two-dimensional ``Coulomb-Dyson gases'', or normal random matrices. The literature on the random matrix models and related Coulomb gases is quite extensive due to the links to various important problems of mathematics and physics.  For an introduction to matrix models see, e.g., \cite{Forrest,Meh}. The normal matrix model was introduced in \cite{CY} in connection to the quantum Hall effect and investigated
further in detail in \cite{CZ}. The connection between the two-dimensional Toda lattice (2DTL) hierarchy and normal random matrices has been established in \cite{CZ}, in \cite{MWZ} it emerged in the context of the
theory of Hele-Shaw flows, or Laplacian growth (for a review see, e.g., \cite{LY_MMNP}).

It is a well-known fact that there is deep connection between the theory of various matrix ensembles and that of tau-function of integrable hierarchies, see, e.g., \cite{KMMOZ,KKMWZ,O} and references therein. Our ``solitonic''
models of statistical mechanics (besides the normal matrix model discussed here, 
one can mention a variety of ensembles
considered in \cite{lou-spi:spectral}) contain partition functions of these ensembles as particular or
limiting cases and therefore generalize this connection: they introduce lattices and 
use non-trivial boundary conditions to the Coulomb-gas models related to the matrix ensembles. 

The Coulomg gas formalism is a powerful tool of treating two-dimensional
problems. Its in depth character is reflected in connecting seemingly different physical notions
like two-dimensional statistical mechanics and quantum field theory models, random matrices 
and integrable systems \cite{Forrest}. It naturally emerges in the investigation of critical
phenomena in two-dimensiona spin lattice systems \cite{N} with further output to
the two-dimensional conformal field theory. In this sense, the interpretation of solitonic
tau-functions as lattice gas (Ising chains) partition functions described below is not an 
occasional coincidence, but yet another indication on the universality of this formalism 
with many physical manifestations.

We start from a brief introduction to basic concepts of statistical mechanics of lattice gases
and its connection to the theory of solitons established previously in
\cite{lou-spi:self,lou-spi:spectral,lou-spi:soliton,lou-spi:critical} (see also, \cite{Za}).
Statistical mechanics studies macroscopic properties of systems with large number of degrees of freedom. Such systems can exist in a discrete (but possibly infinite) set of microstates. A microstate defines the values of all possible microscopic variables.

Consider a lattice consisting of $N$ points (sites) on the plane
which can be occupied by some particles. In the lattice gas model no more than one particle can occupy
each site, i.e. the number of particles at the $i$-th site can be $\nu_i=0$ or $\nu_i=1$. These filling numbers $\nu_1, \nu_2, \dots , \nu_N$ constitute a set of microscopic variables. Then, a microstate of the lattice gas corresponds to a binary string (i.e., a sequence of zeros and ones) of the length $N$. The total number of microstates of the gas equals $2^N$.  Statistical mechanics considers a macroscopic system, that is the $N\to\infty$ ``thermodynamic'' limit, and investigates the corresponding behaviour of the various macroscopic variables,
for example, that of the total number of particles
\begin{equation}
n=\sum_{i=1}^N \nu_i .
\label{nparticles}
\end{equation}
This number $n$ can vary from 0, when the lattice is empty, to $N$ when the lattice is fully occupied by particles.

The probability to find system  in the microscopic state $\nu$ fixed by $N$ filling numbers
$\nu=\{\nu_1, \nu_2, \dots, \nu_N\}$ equals
\begin{equation}
p(\nu)=\frac{1}{Z}e^{-\beta(E(\nu)-\mu n)},
\label{pnu}
\end{equation}
where $E(\nu)=E(\nu_1, \dots, \nu_N)$ is the system energy and $n$ is defined in (\ref{nparticles}). The parameter $\beta>0$ is called the inverse temperature, and the parameter $\mu$ is the chemical potential.
Since the sum of probabilities of all possible micro-states of the system equals to 1, the value $Z$ in the normalization factor  in (\ref{pnu}) equals to the following sum over all possible system states:
\begin{equation}
Z=\sum_{\nu\in {\cal S}}e^{-\beta (E(\nu)-\mu n)}.
\label{Zgrand}
\end{equation}
This sum is called the partition function of the grand canonical ensemble. Here ${\cal S}$ stands for the set of all possible microscopic configurations of the lattice gas given by $2^N$ binary strings $\nu$ of the length $N$.

From definitions (\ref{pnu}) and \eqref{Zgrand}, it follows that
$$
\frac{\partial \log Z}{\partial \beta}=-\langle E \rangle + \mu \langle n \rangle, \quad
\frac{\partial \log Z}{\partial \mu}=\beta \langle n \rangle ,
$$
where
$$
\langle E \rangle=\sum_{\nu\in {\cal S}} p(\nu)E(\nu), \quad \langle n \rangle=\sum_{\nu\in {\cal S}} p(\nu)n(\nu)
$$
are the average energy of the system and the average number of particles in the system respectively.
The quantities $\langle E \rangle$ and $\langle n \rangle$ describe the results of measurements of
the corresponding macroscopic variables.

 One can split the set of microstates of our system ${\cal S}$ into $N+1$ disjoint sets:
 ${\cal S}={\cal S}_0\cup{\cal S}_1\cup{\cal S}_2\cup \dots \cup{\cal S}_N$, where ${\cal S}_i\cap{\cal S}_j=\emptyset$ if $i\not=j$. Here, the set ${\cal S}_0$ corresponds to the states with $n=0$ particles  on the lattice (empty lattice), and the set ${\cal S}_k$ corresponds to the states with $n=k$ with
 ${\cal S}_N$ being the state of fully occupied lattice ($n=N$, i.e. all $\nu_i=1$). Then from (\ref{Zgrand}) we
 have $Z=\sum_{n=0}^N{\cal Z}_ne^{\beta\mu n},$ where
\begin{equation}
{\cal Z}_n=\sum_{\nu\in{\cal S}_n}e^{-\beta E(\nu)}
\label{Zcanonical}
\end{equation}
is the partition function for the system with fixed number of particles $n$. The quantity \eqref{Zcanonical}
is called the partition function of the canonical ensemble, or the $n$-particle partition function.
For the canonical ensemble of $n$ particles we have
$$
\langle E \rangle=-\frac{\partial\log{\cal Z}_n}{\partial \beta}=\frac{\partial\beta F}{\partial \beta} , \quad
F=-\frac{1}{\beta} \log{\cal Z}_n,
$$
where $F$ is called the ``free energy'' of the $n$-particle system. In the thermodynamic limit $n \to\infty$, one usually is interested in the asymptotics of free energy per particle $F/n$.

Let $\zeta_i$ be a complex coordinate of the $i$-th site of the lattice. We consider a system where particles interact pairwise through the two-particle potential $V(z, z')=V(z',z)$ as well as they interact with external fields through a one-particle potential $w(z)$. In other words, the energy of interaction of particles occupying the
$i$-th and $j$-th sites equals $V_{ij}=V(\zeta_i,\zeta_j)$.
We consider only the two-particle interaction with $V(z,z)=+\infty$, so that the condition that the filling factor
$\nu_i$ cannot be greater than 1 will be fulfilled automatically (see below). The energy of interaction of
 $i$-th site's particle with the external fields is $w_i=w(\zeta_i)$. The total energy of the gas then equals to
\begin{equation}
E=\sum_{1\le i<j\le N}V_{ij}\nu_i\nu_j+\sum_{i=1}^N w_i\nu_i.
\label{EnergyNgas}
\end{equation}
This function, relations (\ref{nparticles}) and (\ref{Zgrand}) define the grand partition function of the gas
\begin{equation}
Z=\sum_{\nu_1=0,1}\dots \sum_{\nu_N=0,1} e^{-\beta\left(\sum_{1\le i<j\le N}V_{ij}\nu_i\nu_j+\sum_{i=1}^N (w_i-\mu)\nu_i\right)} .
\label{Latticegas}
\end{equation}
Note that the number $N$ stands here for the number of the lattice sites and not for the number of
particles $n$, which is not fixed in the grand canonical ensemble (one should not confuse $Z$ with
the $N$-particle partition function ${\cal Z}_N$).

For the $n$-particle partition function, defined by (\ref{Zcanonical}) and (\ref{EnergyNgas}), we obtain
\begin{equation}
{\cal Z}_n=\frac{1}{n!}\sum_{z_1\in\zeta}\dots\sum_{z_n\in\zeta}  e^{ -\beta( \sum_{1\le i<j\le n} V(z_i,z_j)+\sum_{i=1}^nw(z_i) )},
\label{Znparticle}
\end{equation}
where $\zeta$ stands for the set of lattice points $\zeta=\{\zeta_1, \zeta_2,\dots ,\zeta_N\}$. Note that, since $V(z,z)=+\infty$, we do not have to care about the configurations in (\ref{Znparticle}) with filling numbers exceeding 1, because all summands with $z_i=z_j$ vanish there. The factor $1/n!$ appears in (\ref{Znparticle}) because, for a given configuration $\nu$ with the number of particles $n$, these $n$ particles are identical in the sum (\ref{EnergyNgas}), while it is not so in (\ref{Znparticle}).

A crucial observation on the connection of lattice statistical mechanics models with the solitons
was done in \cite{lou-spi:self,lou-spi:spectral, lou-spi:soliton}. Namely, soliton $\tau$-functions of integrable hierarchies and partition functions of particular Ising models or lattice gas models have identical structure.
More precisely, the $N$-soliton $\tau$-function of many integrable hierarchies can be written
in the following Hirota form \cite{AS,Hir1976}
\begin{equation}
\tau_N=\sum_{\nu_1=0,1}\dots \sum_{\nu_N=0,1}\exp\left(\sum_{1\le i<j\le N}A_{ij}\nu_i\nu_j+\sum_{i=1}^N\phi_i\nu_i\right),
\label{tauN}
\end{equation}
where the sum is performed over $2^N$ configuration of $N$ discrete variables $\nu_i$, each of $\nu_i$ taking values $0$ or $1$. In equation (\ref{tauN}), $A_{ij}$ stands for the phase shift acquired due to interaction of the $i$-th and $j$-th solitons, while $\phi_i$ stands for the $i$-th soliton phase. The phase shift $A_{ij}$ depends on the momenta of corresponding solitons and the phase $\phi_i$ depends on the momentum of the $i$-th soliton and the full set of the integrable hierarchy times.

In the expression for the $N$-soliton $\tau$-function (\ref{tauN}) one can recognize the grand partition function of a gas models (\ref{Latticegas}) on the lattice consisting of  $N$ sites: the variable $\nu_i$ is the filling factor of the $i$-th lattice site and $\phi_i$ is proportional to the sum of the chemical potential and the external potential at the $i$-th site. In this picture, the phase shift $A_{ij}$ is proportional to the two-body interaction potential between particles occupying the $i$-th and the $j$-th sites:
\begin{equation}
A_{ij}=-\beta V_{ij}, \quad \phi_i=-\beta \left( w_i-\mu \right).
\label{solparV}
\end{equation}
An important point is that this correspondence between the $\tau$-functions and the partition functions
has a restriction --- the parameter $\beta$ appears to be fixed in the $\tau$-functions (see below).

Before proceeding to the main topic, we mention that some connections between soliton solutions of the integrable hierarchies and certain matrix models have been established in \cite{lou-spi:spectral}. In particular, the soliton solutions of the KP hierarchy corresponding to the circular lattice are related to the Gaudin matrix models, which interpolate between the Dyson and uniform unitary ensembles. Also, soliton solutions of the KP and BKP hierarchies corresponding to one-dimensional exponential lattices are related to the one-dimensional translationally-invariant Ising models with non-local interactions. The simplest models of such type are described by the self-similar potentials of the one-dimensional
Schr\"odinger equation \cite{self} and they are related to special infinite-soliton solutions of
the KdV hierarchy \cite{lou-spi:self}. Despite of the fact that the inverse temperature $\beta$ is fixed in
these models, through playing with the model parameters it is possible to go to the zero temperature limit
exhibiting a particular critical behavior of the corresponding Ising chains in the thermodynamic limit
\cite{lou-spi:critical}.

In this paper we consider Coulomb gases on general two-dimensional lattices
and discuss their relation to normal random matrices.
Namely, we derive the partition function of a discretized $n\times n$ normal
matrix model (or a Coulomb-Dyson gas of $n$ charges on $N$ sites on the plane)
from the $N$-soliton $\tau$-function of the two-dimensional Toda lattice (2DTL) hierarchy.

\section{Coulomb-Dyson Gases}

Coulomb gas is a statistical ensemble of charged particles on the plane interacting through the Coulomb potential $V$. On the $(x,y)$-plane the potential created by the particle, placed at the point $(x',y')$, is proportional to the Green function of the 2D Laplace operator
\begin{equation}
\Delta V(x,y; x',y')=-2\pi\delta(x-x')\delta(y-y'), \quad V(x,y; x',y')=V(x',y'; x,y) ,
\label{DeltaG}
\end{equation}
with certain boundary conditions imposed on $V$. For the plane without boundaries
$$
V(z,z')=-\log|z-z'|=-\frac{1}{2}\left(\log(z-z')+\log(\bar z-\bar z')\right),
$$
where the complex variables notation $z=x+\ii y$, $\bar z=x-\ii y$ is used. If the system has an ideal dielectric boundary $\Gamma$, the normal to $\Gamma$ component of the gradient of $V$ vanishes on this boundary, i.e. $(n_x\partial_x+n_y\partial_y)V(x,y,x',y')=0$, where $(n_x,n_y)$ is the normal to $\Gamma$ at the point $(x,y)\in\Gamma$. The tangential component of the gradient vanishes at $\Gamma$, if it is an ideal conductor boundary, i.e. at fixed $z'$ the conductor boundary $\Gamma$ is an equipotential surface of $V$. A useful way of solving (\ref{DeltaG}) with such boundary conditions is provided by the method of images. In what follows, we consider systems where every charge has a finite number of images created by the boundaries.

The electrostatic energy of a system of $n$ particles with charges ${\cal Q}_i$, $i=1,2, \dots, n$, is
\begin{equation}
E_n=\sum_{1\le i<j\le n} {\cal Q}_i{\cal Q}_jV(z_i,z_j)+\sum_{1\le i\le n} {\cal Q}_i^2\tilde V(z_i)
+\sum_{1\le i\le n} {\cal Q}_i W(z_i) ,
\label{energy}
\end{equation}
where $z_i=x_i+\ii y_i$ are the coordinates of the particles. The first term in (\ref{energy}) is the energy of interaction between different charges. The second term is the sum of self-energies: the self-energy of a charge is the energy of interaction between the charge and its own images. The third term describes an interaction of charges with external fields.

The Dyson gas is a one-component (i.e., all particles have equal charges)
Coulomb gas at certain fixed temperatures corresponding to different random matrix models. 
Here, we consider the case $\beta {\cal Q}_i=2$.  Then, without loss of
generality, we can set all ${\cal Q}_i$ in (\ref{energy}) equal to unity and let the inverse
temperature to be $\beta=2$. Next, we consider the lattice version of the Dyson gas
where particles occupy a set of points $\zeta=\{\zeta_1, \dots, \zeta_N\}$ on the $N$ sites lattice
(i.e., $z_i\in\zeta$) and no more than one particle can occupy each site.
Then, the grand partition function of the system corresponding to energy (\ref{energy}) equals
$$
Z=\sum_{\nu_1=0,1}\dots \sum_{\nu_N=0,1}e^{-\beta\left(\frac{1}{2}\sum_{1\le i\not=j\le N}V(\zeta_i,\zeta_j)\nu_i\nu_j+\sum_{i=1}^N (w(\zeta_i)-\mu)\nu_i\right)},
$$
where $w(z)=\tilde V(z)+W(z)$ and $\mu$ stands for the chemical potential.

\section{Ensembles Related to KP, BKP and 2DTL Hierarchies}

It turned out that the Coulomb interaction $V(\zeta_i,\zeta_j)$ can be identified with the phase shifts of different integrable hierarchies \cite{lou-spi:soliton}. In this section we describe connections between the ensembles of the Coulomb-Dyson gases and soliton solutions of the KP (including KdV hierarchy as a subcase)
and the BKP hierarchies established in \cite{lou-spi:soliton} and of the 2DTL hierarchy found in \cite{Za}.

\subsection{KP Hierarchy}

The $N$-soliton $\tau$-function of the Kadomtsev-Petviashvili (KP) hierarchy can be written in the Hirota
form (\ref{tauN}) \cite{Hir1976} with
\begin{equation}
A_{ij}=\log\frac{(a_i-a_j)(b_i-b_j)}{(a_i+b_j)(b_i+a_j)},
\label{shiftsKP}
\end{equation}
$$
\phi_i=\varphi_i+\sum_{n=1}^\infty\left(a_i^n-(-b_i)^n\right)t_n,
$$
where $t_n$, $n=1,2,3, \dots$, is an infinite set of independent variables called the hierarchy ``times'', $(a_i,b_i)$ is the two-dimensional momentum of the $i$-th soliton and $\varphi_i$ is its initial phase. The first non-trivial equation of the hierarchy, the celebrated KP-equation, involves three independent variables $x=t_1$, $y=t_2$ and $t=t_3$ and has the form
\begin{equation}
3\frac{\partial^2u}{\partial y^2}=\frac{\partial}{\partial x}\left(4\frac{\partial u}{\partial t}+6u\frac{\partial u}{\partial x}-\frac{\partial^3 u}{\partial x^3}\right), \quad u=-2\partial^2_x\log\tau .
\label{KPequation}
\end{equation}
All equations of the hierarchy can be encoded in the single bi-linear Hirota residue equation for the $\tau$-function
\begin{equation}
\oint_{z=0}\frac{dz}{2\pi\ii}
e^{\xi(\boldsymbol{t}'-\boldsymbol{t},z)}\tau(\boldsymbol{t}'-[z^{-1}])\tau(\boldsymbol{t}+[z^{-1}])=0,
\label{KPHirota}
\end{equation}
where
\begin{equation}
\boldsymbol{t}=\{t_1,t_2, t_3 \dots\}, \quad \xi(\boldsymbol{t},z)=t_1z+t_2z^2+t_3z^3+\dots, \quad [z^{-1}]=\{z^{-1},\tfrac{1}{2}z^{-2},\tfrac{1}{3}z^{-3}, \dots\},
\label{txiz}
\end{equation}
and $\oint_{z=0}dz/(2\pi\ii)$ (also denoted as $\res_{z=0}$) means taking residue at $z=0$, i.e. a coefficient of $z^{-1}$ in the Laurent series in $z$. A full (infinite) set of the bilinear differential Hirota equations can be obtained from equation (\ref{KPHirota}) as the conditions of vanishing of various coefficients of the Taylor
series in $\varepsilon_i=t'_i-t_i$.

\begin{figure}
\centering
\includegraphics[height=55mm]{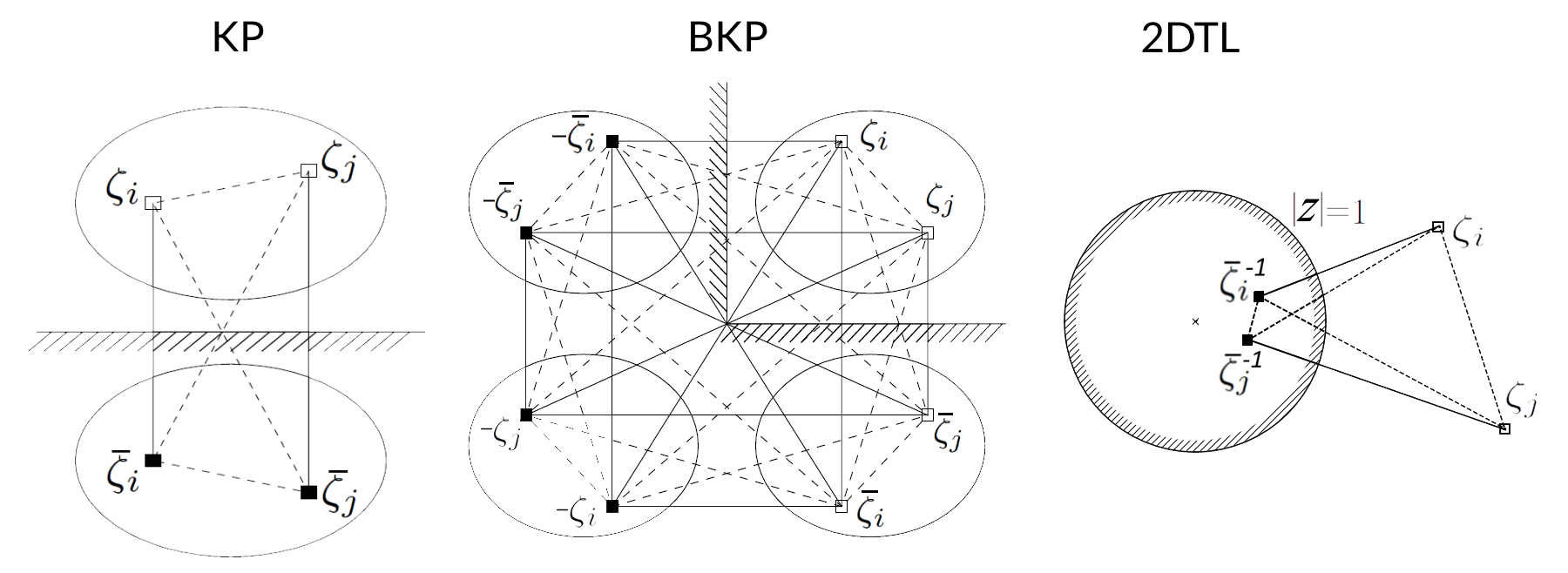}
\caption{From the left to the right. KP hierarchy: a two-dimensional Coulomb gas above an
ideal conductor. BKP hierarchy: a two-dimensional Coulomb gas in the corner between an ideal dielectric (the horizontal axis) and an ideal conductor (the vertical axis). 2DTL hierarchy: a two-dimensional Coulomb gas in the exterior of the unit disc with ideally conducting boundary.  Positive charges are shown as white squares while their negative images are shown as black squares. The interactions between different
charges are shown by dashed lines, while the interactions between
charges and their own images are shown by solid lines.
}
\label{Hierarchies}
\end{figure}

Substituting
\begin{equation}
a_i=\zeta_i, \quad b_i=-\bar \zeta_i, \quad \IM\zeta_i>0, \quad i=1,2, \dots, N,
\label{abz}
\end{equation}
in  (\ref{shiftsKP}), we obtain
$$
V(z,z')=-\log|z-z'|+\log|\bar z-z'| .
$$
This potential satisfies equation (\ref{DeltaG}). For a charge placed at a point $z'$, the real line $\IM z=0$ is an equipotential surface of the potential $V(z,z')$ created by this charge. Therefore, $V(z,z')$ is the Coulomb potential in the upper half-plane with an ideal conducting boundary along the real line.  Equivalently, we may say that this is a potential created by a positive charge at the point $z'$ in the upper half-plane and its reflection image of the opposite charge located in the lower half-plane at $\bar z'$. Thus, the $N$-soliton solution of the KP hierarchy with the momenta defined by (\ref{abz}) and the phases
\begin{equation}
\phi_i=-\beta\left(w(\zeta_i)-\mu\right)=-\beta\left(\log|\bar\zeta_i-\zeta_i|+W(\zeta_i)-\mu\right),
\label{phiplaneKP}
\end{equation}
describes a Coulomb gas at the temperature $\beta=2$ on a lattice in the upper half-plane with the ideally conductive boundary along the $x$-axis (see the first picture on Figure \ref{Hierarchies}).

In (\ref{phiplaneKP}), the potential $w(z)$ is a sum of the self-interaction potential, corresponding to the ``charge-image'' interaction,
$$
\tilde V(z)=\log|\bar z-z|,
$$
and of the external potential $W(z)$. The latter is a sum of a confining potential $U(z)$, keeping particles
in the compact plane domain and determining the initial phases of solitons
$$
\varphi_i=-\beta(U(\zeta_i)-\mu),
$$
and of a harmonic function, determined by the KP ``times'', that corresponds to the electric field created by some distant external charges, i.e.,
$$
W(z)=U(z)-\frac{1}{2}\sum_{p=1}^\infty (z^p-\bar z^p) t_p .
$$
We draw attention to the fact that in order for the external potential to be real all hierarchy ``times''
must be purely imaginary.
The real axis is the equipotential surface of the above harmonic function.

Putting all Coulomb particles on the vertical axis we come to the condition that $a_i=b_i$ for the KP soliton momenta, which corresponds to the KdV-hierarchy solitons.
This system has a natural interpretation as a nonlocal Ising chain with exactly computable
free energy per spin in the translationally invariant cases \cite{lou-spi:self,lou-spi:critical}.

\subsection{BKP Hierarchy}

Soliton solutions of the BKP hietrarchy in the Hirota form (\ref{tauN}) are detrmined by $N$ two-dimensional momenta $(a_i, b_i)$, an infinite number of odd BKP times $t_{2p-1}$, $p=1,2,3, \dots,$ and $N$ initial phases of the solitons $\varphi_i$, so that \cite{HirBKP}
\begin{equation}
A_{ij}=\log\frac{(a_i-a_{j})(b_i-b_{j})
(a_i-b_{j})(b_i-a_{j})}{(a_i+a_{j})(b_i+b_{j})
(a_i+b_{j})(b_i+a_{j})},
\label{shiftsBKP}
\end{equation}
\begin{equation}
\phi_i=\varphi_i+\sum_{p=1}^\infty(a_i^{2p-1}+b_i^{2p-1})t_{2p-1}.
\label{bkp}
\end{equation}
The first non-trivial equation in the hierarchy (the BKP equation) involves three independent variables $t_1$, $t_3$ and $t_5$. Similar to the KP hierarchy \eqref{KPHirota} and \eqref{txiz}, the BKP hierarchy can be encoded in a single Hirota residue equation:
$$
\oint_{z=0}\frac{dz}{2\pi\ii z}
e^{\xi(\boldsymbol{t}'-\boldsymbol{t},z)}\tau(\boldsymbol{t}'-2[z^{-1}])\tau(\boldsymbol{t}+2[z^{-1}])=\tau(\boldsymbol{t})\tau(\boldsymbol{t'}),
$$
with only odd times involved, i.e. $\boldsymbol{t}=\{t_1,t_3,t_5, \dots$\}
and $[z^{-1}]=\{z^{-1},\tfrac{1}{3}z^{-1},\tfrac{1}{5}z^{-5}, \dots\}.$

Choosing soliton momenta as follows
$$
a_i=\zeta_i, \quad b_i=\bar \zeta_i, \quad \RE \zeta_i>0,
$$
the BKP phase shifts (\ref{shiftsBKP}) yield the following two-particle interaction potential
$$
V(z,z')=-\log|z-z'|-\log|\bar z-z'|+\log|z+z'|+\log|\bar z+z'|, \quad \beta=2 ,
$$
which is a Coulomb potential in a quarter of the plane. For convenience, let us choose the right upper corner
of the plane, $\RE z>0$, $\IM z>0$. Then the horizontal boundary of the corner $y=0$ is the ideal dielectric, while the vertical boundary $x=0$ is the ideal conductor. Each charge has now three images: one of the same sign and two of the opposite signs, see picture 2 on Figure \ref{Hierarchies}.

The phases of solitons have the form
$$
\phi_i=-\beta\left(\tilde V(\zeta_i)+W(\zeta_i)-\mu\right),
$$
where the self-interaction potential is
$$
\tilde V(z)=\log|\bar z-z|-\log|\bar z+z|-\log|2z|,
$$
and the external potential $W$ is a sum of the confining potential $U$ and a harmonic function determined by the BKP times (see equations \eqref{bkp} and \eqref{abz})
$$
W(z)=U(z)-\frac{1}{2}\sum_{p=1}^\infty (z^{2p-1}+\bar z^{2p-1})t_{2p-1}.
$$
This harmonic function satisfies all the necessary conditions on the boundaries of the corner.

After setting $a_i=b_i$ in the phase shifts of the BKP hierarchy we obtain the KdV solitons
phase shifts multiplied by 2. This corresponds to the same lattice gas or Ising chain as in the
plain KdV case, but for the twice bigger value of the inverse temperature $\beta=4$ \cite{lou-spi:soliton}.
Again, this value of the temperature corresponds to the particular random matrix ensembles.

\subsection{2TDL Hierarchy}

The two-dimentional Toda lattice (2DTL) hierarchy consists of difference-differential equations \cite{Tak}. It involves an infinite number of continuous independent variables (2DTL times) $t_p$, where $p$ runs over all positive and negative integers, excluding zero, as well as a discrete variable $m\in\mathbb{Z}$. The simplest equation in the hierarchy is the 2DTL equation:
\begin{equation}
\frac{\partial^2u(m)}{\partial t_1 \partial t_{-1}}=e^{-u(m+1)}+e^{-u(m-1)}-2e^{-u(m)},
\label{2DTLe}
\end{equation}
where $u(m)$ depend on $t_p$, $p=\pm 1, \pm 2, \pm 3, \dots $. In terms of the $\tau$-function the 2DTL equation reads as
\begin{equation}
\frac{\partial\tau(m)}{\partial t_1}\frac{\partial\tau(m)}{\partial t_{-1}}-\tau(m)\frac{\partial^2\tau(m)}{\partial t_1 \partial t_{-1}}=\tau(m-1)\tau(m+1), \quad u(m)=\log\frac{\tau(m)^2}{\tau(m+1)\tau(m-1)} .
\label{taum}
\end{equation}

The $N$-soliton $\tau$ function has the form (\ref{tauN}) with the phase shifts
\begin{equation}
A_{ij}=\log\frac{(a_i-a_j)(b_i-b_j)}{(a_i-b_j)(b_i-a_j)},
\label{shifts2DTL}
\end{equation}
which are essentially the KP phase shifts (one should replace $b_i$ to $-b_i$ in (\ref{shiftsKP})), and the soliton phases
$$
\phi_i=\varphi_i+m(\log a_i-\log b_i)+\sum_{p=-\infty, p\not=0}^\infty\left(a_i^p-b_i^p\right)t_p .
$$
Choosing soliton momenta as follows
$$
a_i=\zeta_i/R, \quad b_i=R/\bar\zeta_i, \quad |\zeta_i|>R,
$$
where $R$ is a positive real number, we come to the interaction potential
\begin{equation}
V(z,z')=-\log|z-z'|+\log|\bar zz'-R^2|-\log R, \quad \beta=2 .
\label{VtodaR}
\end{equation}
We set now $R$ equal to 1, $R=1$, and obtain
\begin{equation}
V(z,z')=-\log|z-z'|+\log|\bar zz'-1| .
\label{Vtoda}
\end{equation}
Then $V(z,z')$ satisfies equation (\ref{DeltaG}) in the exterior of the ideally conducting unit disc. In other words, for a charge placed at $z'$, the disc boundary (i.e., the unit circle $|z|=1$) is the equipotential surface of (\ref{Vtoda}). Equivalently, one can say that this is a potential created by a positive charge at the point $z'$, in the exterior of the disc, and its reflection (where reflection is the inversion with respect to the unit circle) image of the opposite charge located inside the disc at $1/\bar z'$, see the third picture on Figure \ref{Hierarchies}.

The self-interaction potential in the 2DTL case is
$$
\tilde V(z)=\log \left|z-\bar z^{-1}\right|,
$$
while the external potential is a sum of the confining potential $U$ and the harmonic function determined by the 2DTL times and the discrete variable $m$:
$$
W(z)=U(z)-m\log|z|-\frac{1}{2}\sum_{p=1}^\infty\left(\left(z^p-\bar z^{-p}\right)t_p+\left(\bar z^p-z^{-p}\right)\bar t_p\right) .
$$
Note, that we have redefined the negative times as $t_{-p}=-\bar t_p$, $p>0$ in order for the potential to be real. Obviously, the unit circle is an equipotential surface of the harmonic field.

Concluding this section we note that the Coulomb interaction potential and the boundary can be conformally transformed. Indeed, let $f(z)$ be a conformal mapping from the exterior of a simple compact domain $\Omega$ to the exterior of the unit circle $|z|=1$, then $\tilde V(z,z')=V(f(z),f(z'))$ is also a Coulomb potential whoose equipotential surface is $\partial\Omega$. Thus $N$-soliton solution with momenta $a_i=f(\zeta_i)$ and $b_i=\bar f(\bar\zeta_i)$ corresponds to the grand partition function of the lattice Coulomb gas in an exterior of conducting domain $\Omega$.

The simplest mapping $f(z)=z/R$ transforms (\ref{Vtoda}) to (\ref{VtodaR}). As an example of less trivial mapping one can take $f(z)=z/2-\sqrt{z^2/4-1}$ (i.e. the mapping inverse to $z\to z+1/z$) from the exterior of the real segment $[-2,2]$ to that of the unit disc. In this case, the ideal conductor is placed along that segment.

\section{2DTL Solitons and Normal Random Matrices}
\label{Rto0}

Let us now remove the constraint $R=1$, and consider a Coulomb gas in the exterior of the disc of an arbitrary radius $R$. Note that the 2DTL hierarchy is invariant under the transformation
$$
m\to m-j, \quad t_p\to R^pt_p, \quad \bar t_p\to R^p\bar t_p, \quad \tau(m) \to R^{m^2+cm+C}\tau(m),
$$
where $c, C$ and $j \in\mathbb{Z}$ are arbitrary constants. In terms of the dependent variable $u$ this transformation produces the shift $u\to u-2\log R$, see (\ref{taum}). In particular, the $\tau$-function
$$
\tilde\tau_N(m, t_1, \bar t_1, t_2, \bar t_2, \dots )=R^{m^2}\tau_N(m-1, Rt_1, R\bar t_1, R^2t_2, R^2\bar t_2, \dots)
$$
is also a solution of the hierarchy. This follows from the bilinear equation (notations are defined in (\ref{txiz}))
$$
\oint_{z=0}\frac{dz}{2\pi\ii}
z^{m'-m}e^{\xi(\boldsymbol{t}'-\boldsymbol{t},z)}\tau(m',\boldsymbol{t}'-[z^{-1}],\boldsymbol{\bar t}')\tau(m,\boldsymbol{t}+[z^{-1}],\boldsymbol{\bar t})=
$$
$$
=\oint_{z=0}\frac{dz}{2\pi\ii}z^{m'-m}e^{\xi(\boldsymbol{\bar t}'-\boldsymbol{\bar t},z^{-1})}\tau(m'+1,\boldsymbol{t}',\boldsymbol{\bar t}'-[z])\tau(m-1,\boldsymbol{t},\boldsymbol{\bar t}+[z])
$$
which encodes the whole hierarchy (for a review see, e.g., \cite{Tak}).

Repeating the above computations  for an arbitrary $R$, we get the transformed $N$-soliton $\tau$-function
\begin{equation}
\tilde\tau_N(m)=R^{m^2}\sum_{\nu_1=0,1}\dots\sum_{\nu_N=0,1}e^{-\beta E(\nu_1, \dots, \nu_N)}, \quad \beta=2,
\label{tauN2DTL}
\end{equation}
where
$$
E=-\sum_{1\leq i<j\leq N}\left(\log|\zeta_i-\zeta_j|-\log|\zeta_i\bar\zeta_j -R^2|+\log R\right)\nu_i\nu_j
$$
$$
-\sum_{i=1}^N \left(\frac{\varphi_i}{2}+(m-1)(\log|\zeta_i|-\log R)+{\cal U}(\zeta_i)\right)\nu_i
$$
and ${\cal U}$ is the harmonic function of the form
$$
{\cal U}(z)=-\frac{1}{2}\sum_{p=1}^\infty \left(\left(z^p-\frac{R^{2p}}{\bar z^p}\right)t_p+\left(\bar z^p-\frac{R^{2p}}{z^p}\right)\bar t_p \right) .
$$

Fixing the initial phases of solitons as
\begin{equation}
\varphi_i=-\beta U(\zeta_i)-\log R, \quad \beta=2,
\label{varphiToda}
\end{equation}
where $U(z)$ is a confining potential, we then take the limit of small $R$. In this limit, when the disc contracts to a point, the harmonic function becomes analytic at $z=0$
\begin{equation}
{\cal U}(z)=-\frac{1}{2}\sum_{p=1}^\infty \left(z^pt_p+\bar z^p\bar t_p \right), \quad R\to 0,
\label{Uharmonic}
\end{equation}
and the energy of the gas becomes
\begin{equation}
E=-\sum_{1\leq i<j\leq N}\nu_i\nu_j\log|\zeta_i-\zeta_j|-\sum_{i=1}^N\left(U(\zeta_i)+{\cal U}(\zeta_i)+(m-n)\log|\zeta_i|\right)\nu_i-n\left(\frac{n}{2}-m\right)\log R.
\label{Etoda}
\end{equation}
Here
$$
n=\sum_{i=1}^N \nu_i
$$
is the number of particles. In the $R\to 0$ limit the last term in (\ref{Etoda}), i.e. $-n\left(\frac{n}{2}-m\right)\log R$, is the dominant one.
Combining it with the factor $R^{m^2}$ in (\ref{tauN2DTL}), we obtain the common factor of
the $n$-particle terms in the $\tilde\tau_N(m)$-function
$$
R^{m^2}e^{\beta n\left(\frac{n}{2}-m\right)\log R}=R^{(m-n)^2},
$$
since $\beta=2$. We see that only the $n$-particle terms with $n=\sum_i \nu_i=m$ are finite for $R\to 0$, while other terms vanish. Also, since the total number of gas particles $n$ cannot be negative or exeed the number of lattice sites $N$, the lowest vanishing rate of $\tilde\tau_N(m)$  for $m\le 0$ is reached at $n=0$, while for $m>N$ it is reached at $n=N$ (i.e., at one of the ends of the interval $0\le n\le N$), so that all the terms in (\ref{tauN2DTL}) for $m$ outside of this interval vanish. Thus, in the limit $R\to 0$, we obtain
$$
\tilde\tau_N(m)={\cal Z}_m, \quad 0\le m\le N , \quad \tilde\tau_N(m<0)=\tilde\tau_N(m>N)=0,
$$
where ${\cal Z}_m$ is the partition function of the gas of $m$ Coulomb particles on the lattice $\zeta=\{\zeta_1, \zeta_2, \dots , \zeta_N\}$ in the complex plane without boundaries. No more than one particle can occupy a lattice site and
\begin{equation}
{\cal Z}_m=\frac{1}{m!}\sum_{z_1\in\zeta}\dots\sum_{z_m\in\zeta}e^{-\beta E_m(z_1, \dots, z_m)} , \quad \beta=2 ,
\label{Zn2DTL}
\end{equation}
with the gas energy being
\begin{equation}
E_m(z_1, \dots, z_m)=-\sum_{1\le i<j\le m}\log|z_i-z_j|+\sum_{1\le i\le m}\left(U(z_i)+{\cal U}(z_i)\right).
\label{En}
\end{equation}
Finally, we can write ${\cal Z}_0=1$ and
\begin{equation}
{\cal Z}_m=\frac{1}{m!}\sum_{\ell_1=1}^N\dots\sum_{\ell_m=1}^N \prod_{1\le i< j\le m}
\left|\zeta_{\ell_i}-\zeta_{\ell_j}\right|^2\prod_{i=1}^m \exp\left(-2U(\zeta_{\ell_i})-2{\cal U}(\zeta_{\ell_i})\right), \quad m>0.
\label{Zn2DTLprod}
\end{equation}

Thus, we have shown that the $m$-particle partition function ${\cal Z}_m$, determined in \eqref{Zn2DTL},
\eqref{En} or (\ref{Zn2DTLprod}), with the harmonic external potential ${\cal U}$ fixed in (\ref{Uharmonic}),
is a $\tau$-function of the two-dimensional Toda lattice of the length $N-1$. In other words, the function
\begin{equation}
\tau(m, t_1, \bar t_1, t_2, \bar t_2, \dots)=\left\{
\begin{array}{c}
{\cal Z}_{m}(t_1, \bar t_1, t_2, \bar t_2, \dots), \quad {\rm if} \quad m=0,1,2, \dots, N, \\
0, \quad {\rm otherwise,}
\end{array}\right.
\label{tauZm}
\end{equation}
is a solution of the 2DTL hierarchy. In terms of $u(m)$, the 2DTL equation (\ref{2DTLe}) becomes a system of $N-1$ differential equations for $u(1), \dots , u(N-1)$. For $m=2,3, \ldots, N-2$, the equations are fixed in (\ref{2DTLe}), while for $m=1$ and $m=N-1$ we have $\partial_{t_1}\partial_{t_{-1}}u(1)=e^{-u(2)}-2e^{-u(1)}$
and $\partial_{t_1}\partial_{t_{-1}}u(N-1)=e^{-u(N-2)}-2e^{-u(N-1)}$, respectively.

The $m$-particle partition function (\ref{Zn2DTLprod}) (or, equivalently, \eqref{Zn2DTL}, \eqref{En})
can be rewritten in the form
\begin{equation}
{\cal Z}_m=\frac{1}{m!}\int \prod_{1\le i<j\le m} |z_i-z_j|^2 \prod_{j=1}^m e^{-2{\cal U}(z_j)}\rho(z_j)dx_jdy_j,
\label{normal}
\end{equation}
where
$$
\rho(z)=\sum_{i=1}^N e^{-2U(z)}\delta(x-X_i)\delta(y-Y_i), \quad \zeta_i=X_i+\ii Y_i .
$$
In the continuous $N\to\infty$ limit, the spacing between sites tends to zero, i.e. $\zeta_i=\epsilon\xi_i$,
$\epsilon\to 0$, while the area of the lattice, which is of the
order of $\epsilon^2 N$, remains finite. In this limit
the measure $\rho$ tends to the continuous measure
\begin{equation}
\rho(z)=\epsilon^{-2} e^{-2U(z)}\varrho(z),
\label{rho}
\end{equation}
where $\varrho(z)$ is a normalized finite density of the lattice sites. We recall that the $\tau$-function is defined modulo a gauge factor, and, in particular, it can be multiplied by any constant. Therefore, the diverging factor $\epsilon^{-2}$ can be discarded in the above equality.

The $\tau$-function (\ref{normal}) is the partition function of the random normal $m\times m$ matrix model
Depending on the measure, we obtain either continuous or discrete version of the model.
Note that the standard hermitian matrix model is a special case of the normal matrix model,
when the measure $\rho$ is concentrated on the real line, i.e. this model enters our considerations as well.

By choosing $\varphi_i=-\beta U(\zeta_i)+(2\ell-1)\log R$, $\ell\in \mathbb{Z}$, instead of (\ref{varphiToda}),
we will get a partition function of the Coulomb gas in the presence of a fixed point charge of the value
$\ell$ placed at the point $z=0$. Determinantal representations of the partition functions of gases
with several fixed charges were given, e.g., in \cite{CZ,HO}.

Note that the partition function (\ref{normal}) of the normal matrix model with an arbitrary (continuous or/and discrete) measure can be obtained without the lattice scaling procedure, by applying our limiting procedure directly to the well-known solution of the hierarchy (see e.g. \cite{JimMiw,Za})
$$
\tau=\sum_{n\ge 0}\frac{1}{n!}\int\dots\int\prod_{1\le i<j\le n}\frac{(p_i-p_j)(q_i-q_j)}{(p_i-q_j)(q_i-p_j)}\prod_{i=1}^n e^{m(\log p_i-\log q_i)
+\sum_{k=-\infty \atop \makebox[-1em]{} k\not=0}^\infty\left(p_i^k-q_i^k\right)t_k} r(p_i,q_i)dp_idq_i,
$$
for which the $N$-soliton solution is a particular case corresponding to $r(p,q)=\sum_{i=1}^N r_i\delta(p-a_i)(q-b_i)$. The procedure is essentially the same as in the case of the $N$-soliton solutions, except the continuum limit at the final stage is not needed.

\section{Conclusions}

The grand partition function of the Coulomb-Dyson gas and the $m$-particle partition function related to the normal matrix model are usually obtained by using different ``group-like'' elements in the free-fermion approach to integrable hierarchies \cite{Za}. In this article, we have shown that the latter can be obtained from the former one by a special limiting procedure.

It is well known that the partition function of the standard normal random matrix model
having the continuous measure emerges in the dispersionless limit of the 2DTL
hierarchy corresponding to a special $m \to \infty$ scaling limit. In the dispersionless limit one introduces
the scaled (``slow'') times $T_i$, such that $T_{\pm i}=ht_{\pm i}$, and $T_0=hm$. Here, one considers
the double scaling limit, such that $h\to 0$, $m\to\infty$ with $T_0$ remaining finite.
The scaling parameter $h$ plays the role of spacing on the Toda lattice (i.e., the minimal step
in ``time'' $T_0$, not to confuse with spacing on the soliton momentum lattice).
When $h\to 0$, the time $T_0$ becomes the continuous variable
for slowly varying solutions. For instance, in this limit the Toda lattice equation (\ref{2DTLe}) becomes
$$
\frac{\partial^2 u}{\partial T_1\partial{T_{-1}}}=\frac{\partial^2 e^{-u}}{\partial T_0^2}.
$$
Commutators in the Lax representation of the hierarchy degenerate to the Poisson brackets and the Lax operators become functions, i.e. the dispersionless limit is a kind of a quasiclassical limit with $h$ playing a role of the ``Plank constant'' (for more details see e.g. \cite{TakTak}).

From the point of view of the matrix models, introduction of the ``slow'' times corresponds to the scaling of the harmonic potential ${\cal U}(z,t)={\cal U}(z,T)/h$ (see eq. (\ref{Uharmonic})). Also by making the measure (\ref{rho}) dependent on the parameter $h$ through the $h$-dependence of the confining potential $U(z,h)=v(z)/h$, we get the matrix model with the scaled total potential $U(z,h)+{\cal U}(z,t)=\frac{1}{h}\left(v(z)+{\cal U}(z,T)\right)$.  The potential diverges as $h\to 0$, while $hm$  (where $m$ is the size of matrix) remains finite. In this limit, for a wide class of potentials, the eigenvalues of random normal matrices occupy a compact
domain in the complex plane, called a ``droplet''.
In the case of the confining potential created by the uniform neutralizing background, the evolution of
boundaries of this droplet with respect to the matrix size, i.e. with respect to the time $T_0$,
is a solution to the Laplacian growth, or Hele-Shaw moving boundary problem of the 2D fluid dynamics (with the area of the droplet proportional to $T_0$, see, e.g., \cite{HLY,LY_MMNP,MPT, MWZ}).

The Laplacian growth is a model of evolution of
a droplet whose boundary is driven by a harmonic field being a potential for the
growth velocity. The field is a Green function of the exterior of the droplet which vanishes on the droplet boundary. The normal velocity of the boundary is proportional to the normal derivative of the field.
The problem in which the droplet is expanding is linearly unstable and
ill-posed for almost any initial condition (see, e.g. \cite{GTV}
and references therein). The finite $m$ normal matrix model provides a sort of integrable regulaization of the above ill-posed problem. The discretization of the matrix model, related to $N$-soliton solutions, could provide another type of the regularization, a kind of the lattice regularization. In this case, the continuous Laplacian growth is recovered in a pair of double scaling limits $N\to\infty, \epsilon\to 0$, $m\to\infty, h\to 0$.

It is worth mentioning that apart from the standard normal matrix model, which is a generalization of the hermitian random matrix model to normal matrices, non-hermitian generalizations of the symmetric and quarternion-real models were considered in the literature as well, see \cite{O} and references therein. These models, called ``generalized Ginibre ensembles'', turn out to be related to the so-called ``large BKP'' and ``large 2-BKP" hierarchies. It would be interesting to study solitonic $\tau$-functions of these
hierarchies in the context of the statistical mechanics of Coulomb gases with boundaries. Some soliton solutions of the large (``fermionic") BKP hierarchy were written down in \cite{OST} and this may be useful for further studies. 

It can be also interesting to try to apply our method to the Mehta-Pandey interpolating ensembles,
to circular $\beta=1,4$ ensembles \cite{Meh} and also to certain new solvable ensembles of
random matrices such as recently considered in \cite{AOV} and \cite{O1}.

\begin{figure}
\centering
\includegraphics[height=55mm]{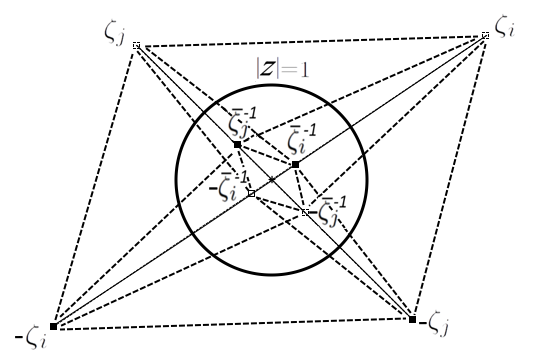}
\caption{Positive charges/images are shown as white squares while negative images are shown as black squares. The interactions between different
charges are shown by dashed lines, while the interactions between
charges and their own images are shown by solid lines.
}
\label{BKP2}
\end{figure}

Note that since the KP and 2DTL phase shifts are essentialy the same, the Coulomb potential related to the 2DTL could be already obtained from the KP phase shifts (\ref{shiftsKP}) by setting $a_i=\zeta_i/R$, $b_i=-R/\zeta_i$. One might try to apply this choice to the BKP phase shifts (\ref{shiftsBKP}). However, the corresponding interaction potential has no clear physical meaning in this case:
here not only images which are reflections $\zeta_i \to R^2/\bar \zeta_i$ with
respect to the circle, but also the inversions $\zeta_i\to-\zeta_i$ with respect to the
origin are present (see Figure \ref{BKP2}). 

Concluding this article we would like to remind that, while useful from the point of view of random matrices and
Dyson gases, the $\tau$-function approach has an essential drawback in the framework of general statistical mechanics. Namely, the partition functions derived from the $\tau$-functions of integrable hierarchies correspond to the Coulomb gas (or Ising models \cite{lou-spi:self,lou-spi:spectral}) models at fixed (inverse) temperatures $\beta=2$. For the translationally invariant Ising models with non-local interaction in one dimension,
related to the self-similar potentials \cite {self} and certain soliton solutions of the KdV and BKP hierarchies \cite{lou-spi:self,lou-spi:spectral}, we have temperatures $\beta=2$ and $\beta=4$, respectively.
Restriction to fixed temperatures is a consequence of the fact that the integrable hierarchies are nothing, but the Plucker relations on an infinite dimensional Grassmanian \cite{DatJimKashMiw,JimMiw,Sato} that can be obtained in the framework of the free-fermion formalism.

\smallskip

{\bf Acknowledgments.} This study has been partially funded within the framework
of the HSE University Basic Research Program. I. Loutsenko and O. Yermolaeva would like to thank
Centre de Recherches Math\'ematiques for a support. The authors thank also the referees for their
helpful remarks.

\end{document}